\documentstyle[overcite]{article}

\begin{document}


\newenvironment{changemargin}[2]{\begin{list}{}{
   \setlength{\topsep}{0pt}
   \setlength{\leftmargin}{0pt}
   \setlength{\rightmargin}{0pt}
   \setlength{\listparindent}{\parindent}
   \setlength{\itemindent}{\parindent}
   \setlength{\parsep}{0pt plus 1pt}
   \addtolength{\leftmargin}{#1}\addtolength{\rightmargin}{#2}
   }\item }{\end{list}}

%
\newdimen\myfloatwidth
\newenvironment{RHStable}[1]
{
\begin{table}[hbt]
\myfloatwidth=\textwidth
\advance\myfloatwidth by -#1
\begin{changemargin}{\myfloatwidth}{0cm}
\centering
}
{
\end{changemargin}
\end{table}
}
\newenvironment{LHStext}[2]
{
\vspace{-#2}
\myfloatwidth=#1
\advance\myfloatwidth by 0.5cm
\begin{changemargin}{0cm}{\myfloatwidth}
}
{
\end{changemargin}
}

\newcommand{\magnitude}[2]{\mbox{$ #1 \!\! \stackrel{m}{.} \!\! #2 $}}

\newcommand{\dotdeg}[2]{\mbox{$ #1 \! \stackrel{\rm o}{\vspace{3mm} .} \! #2 $}}

\newcommand{\degrees}[1]{\mbox{$ #1 ^{\circ} $}}

\newcommand{\real}[2]{\mbox{$ #1 {\rm x} 10^{#2} $}}

\newcommand{\err}[1]{\mbox{$\pm$}#1}

\newcommand{\Stromgren}{Str{\"{o}}mgren}
\newcommand{\Angstrom}{{\AA}ngstr{\"{o}}m}
\newcommand{\Cortes}{Cort{\'{e}}s}
\newcommand{\Gutierrez}{Guti{\'{e}}rrez-Moreno}
\newcommand{\Tug}{T{\"{u}}g}
\newcommand{\Bartkevicius}{Bartkevi{\u{c}}ius}
\newcommand{\Bridzius}{Brid{\u{z}}ius}
\newcommand{\Burnasov}{Burna{\u{s}}ov}
\newcommand{\Cernies}{{\u{C}}ernies}
\newcommand{\Dzervitis}{Dz{\'{e}}rv{\'{\i}}tis}
\newcommand{\Jasevicius}{Jasevi{\u{c}}ius}
\newcommand{\Kaccinskas}{Ku{\u{c}}inskas}
\newcommand{\Kuriliene}{Kurilien{\.{e}}}
\newcommand{\Lazauskaite}{Lazauskait{\.{e}}}
\newcommand{\Meistas}{Mei{\u{s}}tas}
\newcommand{\Sleivyte}{{\u{S}}leivyt{\.{e}}}
\newcommand{\Straizys}{Strai{\u{z}}ys}
\newcommand{\Sudzius}{S{\={u}}d{\u{z}}ius}
\newcommand{\Sviderskiene}{Sviderskien{\.{e}}}
\newcommand{\Tautvaisiene}{Tautvai{\u{s}}ien{\.{e}}}
\newcommand{\Vansevicius}{Vansevi{\u{c}}ius}
\newcommand{\Zdanavicius}{Zdanavi{\u{c}}ius}

\newcommand{\John}{Johnson}

\newcommand{\StdAirmassPlot}{Bouguer}

\newcommand{\McLellan}{${\rm M^cLellan}$}

\newcommand{\KCrux}{\mbox{{\normalsize $\!\!\!\kappa$}~Crucis}} 
\newcommand{\KCrucis}{\mbox{{\Large $\!\!\!\kappa$}~Crucis}} 
\newcommand{\OVel}{\mbox{O~Vel}}
\newcommand{\RAra}{\mbox{R~Arae}}

\newcommand{\approxgreater}{\mbox{\scriptsize $\widetilde{>}$}}
\newcommand{\approxlesser}{\mbox{\scriptsize $\widetilde{<}$}}
\newcommand{\approxGreater}{\mbox{\scriptsize $ \stackrel{>}{\scriptstyle \sim} $}}
\newcommand{\approxLesser}{\mbox{\scriptsize $ \stackrel{<}{\scriptstyle \sim} $}}

\newcommand{\Hb}{\mbox{$H_{\beta}$}}

\newcommand{\km}{\mbox{$km$}}
\newcommand{\mm}{\mbox{$mm$}}
\newcommand{\cm}{\mbox{$cm$}}
\newcommand{\nm}{\mbox{$nm$}}
\newcommand{\um}{\mbox{$\mu m$}}
\newcommand{\Ang}{\mbox{\AA}}
\newcommand{\Torr}{\mbox{$Torr$}}
\newcommand{\Hg}{\mbox{$Hg$}}
\newcommand{\atm}{\mbox{$atm$}}
\newcommand{\atmcm}{\mbox{$atm\!-\!cm$}}
\newcommand{\Kelvin}{\mbox{$Kelvin$}}
\newcommand{\Celsius}{\mbox{$C$}}
\newcommand{\Volts}{\mbox{$V$}}


\def\diameter{{\ifmmode\mathchoice
{\ooalign{\hfil\hbox{$\displaystyle/$}\hfil\crcr
{\hbox{$\displaystyle\mathchar"20D$}}}}
{\ooalign{\hfil\hbox{$\textstyle/$}\hfil\crcr
{\hbox{$\textstyle\mathchar"20D$}}}}
{\ooalign{\hfil\hbox{$\scriptstyle/$}\hfil\crcr
{\hbox{$\scriptstyle\mathchar"20D$}}}}
{\ooalign{\hfil\hbox{$\scriptscriptstyle/$}\hfil\crcr
{\hbox{$\scriptscriptstyle\mathchar"20D$}}}}
\else{\ooalign{\hfil/\hfil\crcr\mathhexbox20D}}%
\fi}}
\def\sun{\hbox{$\odot$}}
\def\degr{\hbox{$^\circ$}}
\def\arcmin{\hbox{$^\prime$}}
\def\arcsec{\hbox{$^{\prime\prime}$}}
\def\utw{\smash{\rlap{\lower5pt\hbox{$\sim$}}}}
\def\udtw{\smash{\rlap{\lower6pt\hbox{$\approx$}}}}
\def\fd{\hbox{$.\!\!^{\rm d}$}}
\def\fh{\hbox{$.\!\!^{\rm h}$}}
\def\fm{\hbox{$.\!\!^{\rm m}$}}
\def\fs{\hbox{$.\!\!^{\rm s}$}}
\def\fdg{\hbox{$.\!\!^\circ$}}
\def\farcm{\hbox{$.\mkern-4mu^\prime$}}
\def\farcs{\hbox{$.\!\!^{\prime\prime}$}}
\def\fp{\hbox{$.\!\!^{\scriptscriptstyle\rm p}$}}

%
%

\def\la{\mathrel{\mathchoice {\vcenter{\offinterlineskip\halign{\hfil
$\displaystyle##$\hfil\cr<\cr\sim\cr}}}
{\vcenter{\offinterlineskip\halign{\hfil$\textstyle##$\hfil\cr
<\cr\sim\cr}}}
{\vcenter{\offinterlineskip\halign{\hfil$\scriptstyle##$\hfil\cr
<\cr\sim\cr}}}
{\vcenter{\offinterlineskip\halign{\hfil$\scriptscriptstyle##$\hfil\cr
<\cr\sim\cr}}}}}
\def\ga{\mathrel{\mathchoice {\vcenter{\offinterlineskip\halign{\hfil
$\displaystyle##$\hfil\cr>\cr\sim\cr}}}
{\vcenter{\offinterlineskip\halign{\hfil$\textstyle##$\hfil\cr
>\cr\sim\cr}}}
{\vcenter{\offinterlineskip\halign{\hfil$\scriptstyle##$\hfil\cr
>\cr\sim\cr}}}
{\vcenter{\offinterlineskip\halign{\hfil$\scriptscriptstyle##$\hfil\cr
>\cr\sim\cr}}}}}
\def\cor{\mathrel{\mathchoice {\hbox{$\widehat=$}}{\hbox{$\widehat=$}}
{\hbox{$\scriptstyle\hat=$}}
{\hbox{$\scriptscriptstyle\hat=$}}}}
\def\sol{\mathrel{\mathchoice {\vcenter{\offinterlineskip\halign{\hfil
$\displaystyle##$\hfil\cr\sim\cr<\cr}}}
{\vcenter{\offinterlineskip\halign{\hfil$\textstyle##$\hfil\cr\sim\cr
<\cr}}}
{\vcenter{\offinterlineskip\halign{\hfil$\scriptstyle##$\hfil\cr\sim\cr
<\cr}}}
{\vcenter{\offinterlineskip\halign{\hfil$\scriptscriptstyle##$\hfil\cr
\sim\cr<\cr}}}}}
\def\sog{\mathrel{\mathchoice {\vcenter{\offinterlineskip\halign{\hfil
$\displaystyle##$\hfil\cr\sim\cr>\cr}}}
{\vcenter{\offinterlineskip\halign{\hfil$\textstyle##$\hfil\cr\sim\cr
>\cr}}}
{\vcenter{\offinterlineskip\halign{\hfil$\scriptstyle##$\hfil\cr
\sim\cr>\cr}}}
{\vcenter{\offinterlineskip\halign{\hfil$\scriptscriptstyle##$\hfil\cr
\sim\cr>\cr}}}}}
\def\lse{\mathrel{\mathchoice {\vcenter{\offinterlineskip\halign{\hfil
$\displaystyle##$\hfil\cr<\cr\simeq\cr}}}
{\vcenter{\offinterlineskip\halign{\hfil$\textstyle##$\hfil\cr
<\cr\simeq\cr}}}
{\vcenter{\offinterlineskip\halign{\hfil$\scriptstyle##$\hfil\cr
<\cr\simeq\cr}}}
{\vcenter{\offinterlineskip\halign{\hfil$\scriptscriptstyle##$\hfil\cr
<\cr\simeq\cr}}}}}
\def\gse{\mathrel{\mathchoice {\vcenter{\offinterlineskip\halign{\hfil
$\displaystyle##$\hfil\cr>\cr\simeq\cr}}}
{\vcenter{\offinterlineskip\halign{\hfil$\textstyle##$\hfil\cr
>\cr\simeq\cr}}}
{\vcenter{\offinterlineskip\halign{\hfil$\scriptstyle##$\hfil\cr
>\cr\simeq\cr}}}
{\vcenter{\offinterlineskip\halign{\hfil$\scriptscriptstyle##$\hfil\cr
>\cr\simeq\cr}}}}}
\def\grole{\mathrel{\mathchoice {\vcenter{\offinterlineskip\halign{\hfil
$\displaystyle##$\hfil\cr>\cr\noalign{\vskip-1.5pt}<\cr}}}
{\vcenter{\offinterlineskip\halign{\hfil$\textstyle##$\hfil\cr
>\cr\noalign{\vskip-1.5pt}<\cr}}}
{\vcenter{\offinterlineskip\halign{\hfil$\scriptstyle##$\hfil\cr
>\cr\noalign{\vskip-1pt}<\cr}}}
{\vcenter{\offinterlineskip\halign{\hfil$\scriptscriptstyle##$\hfil\cr
>\cr\noalign{\vskip-0.5pt}<\cr}}}}}
\def\leogr{\mathrel{\mathchoice {\vcenter{\offinterlineskip\halign{\hfil
$\displaystyle##$\hfil\cr<\cr\noalign{\vskip-1.5pt}>\cr}}}
{\vcenter{\offinterlineskip\halign{\hfil$\textstyle##$\hfil\cr
<\cr\noalign{\vskip-1.5pt}>\cr}}}
{\vcenter{\offinterlineskip\halign{\hfil$\scriptstyle##$\hfil\cr
<\cr\noalign{\vskip-1pt}>\cr}}}
{\vcenter{\offinterlineskip\halign{\hfil$\scriptscriptstyle##$\hfil\cr
<\cr\noalign{\vskip-0.5pt}>\cr}}}}}
\def\loa{\mathrel{\mathchoice {\vcenter{\offinterlineskip\halign{\hfil
$\displaystyle##$\hfil\cr<\cr\approx\cr}}}
{\vcenter{\offinterlineskip\halign{\hfil$\textstyle##$\hfil\cr
<\cr\approx\cr}}}
{\vcenter{\offinterlineskip\halign{\hfil$\scriptstyle##$\hfil\cr
<\cr\approx\cr}}}
{\vcenter{\offinterlineskip\halign{\hfil$\scriptscriptstyle##$\hfil\cr
<\cr\approx\cr}}}}}
\def\goa{\mathrel{\mathchoice {\vcenter{\offinterlineskip\halign{\hfil
$\displaystyle##$\hfil\cr>\cr\approx\cr}}}
{\vcenter{\offinterlineskip\halign{\hfil$\textstyle##$\hfil\cr
>\cr\approx\cr}}}
{\vcenter{\offinterlineskip\halign{\hfil$\scriptstyle##$\hfil\cr
>\cr\approx\cr}}}
{\vcenter{\offinterlineskip\halign{\hfil$\scriptscriptstyle##$\hfil\cr
>\cr\approx\cr}}}}}
\def\lid{\mathrel{\mathchoice {\vcenter{\offinterlineskip\halign{\hfil
$\displaystyle##$\hfil\cr<\cr\noalign{\vskip1.2pt}=\cr}}}
{\vcenter{\offinterlineskip\halign{\hfil$\textstyle##$\hfil\cr<\cr
\noalign{\vskip1.2pt}=\cr}}}
{\vcenter{\offinterlineskip\halign{\hfil$\scriptstyle##$\hfil\cr<\cr
\noalign{\vskip1pt}=\cr}}}
{\vcenter{\offinterlineskip\halign{\hfil$\scriptscriptstyle##$\hfil\cr
<\cr
\noalign{\vskip0.9pt}=\cr}}}}}
\def\gid{\mathrel{\mathchoice {\vcenter{\offinterlineskip\halign{\hfil
$\displaystyle##$\hfil\cr>\cr\noalign{\vskip1.2pt}=\cr}}}
{\vcenter{\offinterlineskip\halign{\hfil$\textstyle##$\hfil\cr>\cr
\noalign{\vskip1.2pt}=\cr}}}
{\vcenter{\offinterlineskip\halign{\hfil$\scriptstyle##$\hfil\cr>\cr
\noalign{\vskip1pt}=\cr}}}
{\vcenter{\offinterlineskip\halign{\hfil$\scriptscriptstyle##$\hfil\cr
>\cr
\noalign{\vskip0.9pt}=\cr}}}}}

\newcommand{\MyNewLine}{\vspace{3mm}}

\begin{center}
{\large NOTES FROM OBSERVATORIES}\\
\MyNewLine
MOUNT PINATUBO AND
ATMOSPHERIC EXTINCTION AT MOUNT~JOHN UNIVERSITY OBSERVATORY 1987-94\\
\MyNewLine
{\it
By M.\ C.\ Forbes, T.\ Banks \& D.\ J.\ Sullivan\\
Physics Dept., Victoria University of Wellington, New Zealand\\
\MyNewLine
and\\
\MyNewLine
R.\ J.\ Dodd\\
Carter Observatory, Wellington, New Zealand\\
\MyNewLine
and\\
\MyNewLine
A.\ C.\ Gilmore \& P.\ M.\ Kilmartin\\
Mt.\ John University Observatory, Canterbury University, New Zealand\\
\MyNewLine}
\end{center}

A variety of photometric observing programmes undertaken at Mt.\ John
University Observatory (MJUO) obtain atmospheric extinction data in several
passbands.  The atmospheric extinction in the V passband measured at MJUO
from late 1987 to early 1994 is shown in Fig.\ 1.  Every data point is the
average extinction over each night observed and has been derived from one
of several instrumental systems and reduction algorithms.

The first system is a \John\ V filter on either a Boller \& Chivens or an
Optical Craftsman 60cm telescope, with the extinction derived from a
\StdAirmassPlot\ plot\cite{kn:Hardie} using the programs of Henden and
Kaitchuck
\cite{kn:HendonKaitchuckPC}.

The next system is a Vilnius V filter on the same telescopes.  This is a
medium pass-band system designed to allow spectral classification of stars
to be made without recourse to spectroscopy\cite{kn:StraizysBook}.  The
mean wavelength of the response function of the Vilnius V filter nearly
coincides with that of the \John\ V, allowing the V extinction between the
two systems to be considered the same (to an first order approximation).  A
modified version\cite{kn:BalticPaperI} of Nikonov's
method\cite{kn:Nikonov1953,kn:StraizysBook} was used to measure the
extinction once every hour during the night.

The third system is a \John\ filter set on the \McLellan\ one metre
telescope using a Thomson CCD with a Photometrics control
system\cite{kn:TobinCCD}.  The extinction was calculated from observations
of standard stars over a wide range of airmass, simultaneously fitting the
standardisation transformation and extinction coefficients as described by
Harris {\it et~al.\ }\cite{kn:HarrisReduction} and implemented in the
Image Analysis and Reduction Facility (IRAF)\cite{kn:TodyIRAF}.

The final system is a two channel photomultiplier instrument on the one
metre telescope using a \John\ V filter\cite{kn:Sullivan2Channel}.  The
extinction data were derived from custom written software based on a
\StdAirmassPlot\ plot.

The extinction values found by the different systems were compared for
common nights and agree within 0.03 mags/airmass.  As another check, the
average extinction values (before and after the Mt.\ Pinatubo eruption) of
each system were calculated and again all found to agree within 0.03
mags/airmass.

The effect of the 15 June 1991 Mt.\ Pinatubo eruption can clearly be seen
in Fig.\ 1, with the dashed lines representing the average extinction
before and after the eruption.  This was the only volcanic event during the
period 1987-94 to have a significant effect on the
stratosphere\cite{kn:SpecialIssue}.  The increase in average V extinction
(0.08~mags/airmass) in similar to that found at La Silla
(0.078~mags/airmass), SAAO (0.06~mags/airmass) and Kitt Peak
(0.08~mags/airmass)\cite{kn:Grothues}.

\begin{figure}
\vspace{55mm}
{\footnotesize
\begin{center}
Fig.\ 1\\
\end{center}
\begin{changemargin}{6mm}{6mm}
The V passband extinction at Mount John University Observatory over 1987 -
94, with data combined from four different instrumental systems and
reduction algorithms.  The dashed lines represent the average values before
and after the Mt.\ Pinatubo eruption, the solid line is a running average.
\end{changemargin}
}
\end{figure}

It took approximately 80 days for the aerosol cloud to reach New Zealand
\cite{kn:WadsworthBlackBirch}, with the extinction starting to increase
around JD~2448500, and reaching an initial maximum about 1 October 1991
(JD~2448530).  As of June 1994, the extinction has not yet returned to
pre-eruption levels.  Further, the night-to-night variations hide any
exponential decay back towards the Mt.\ Pinatubo pre-eruption levels, as
also found at La Silla\cite{kn:Grothues}.

The data were also investigated for seasonal effects, as there appeared to
be an increase in the extinction every July since the eruption.  A sinusoid
was fitted\cite{kn:Rufener}, but was not found to be statistically more
significant than a single (mean) extinction value.  Hence the
night-to-night variations in extinction are greater than any seasonal
changes, emphasizing that extinction should be determined for each night
rather than using monthly averages.

\newpage
\MyNewLine\noindent{\it Acknowledgements}
\MyNewLine

MF, TB, DJS and RJD would like to thank University of Canterbury and Mt.\
John University Observatory staff for the generous use of their facilities,
the New Zealand Foundation for Research, Science and Technology and the VUW
Internal Research Grant Committee for partial funding of their projects.
It is a pleasure to thank Prof.\ V.\ \Straizys\ and the Lithuanian
Institute of Theoretical Physics and Astronomy for supplying the Vilnius
filter sets.  TB acknowledges partial support during this study by the
inaugural Royal Society of New Zealand R.H.T Bates Postgraduate
Scholarship. 

\bibliographystyle{unsrt}

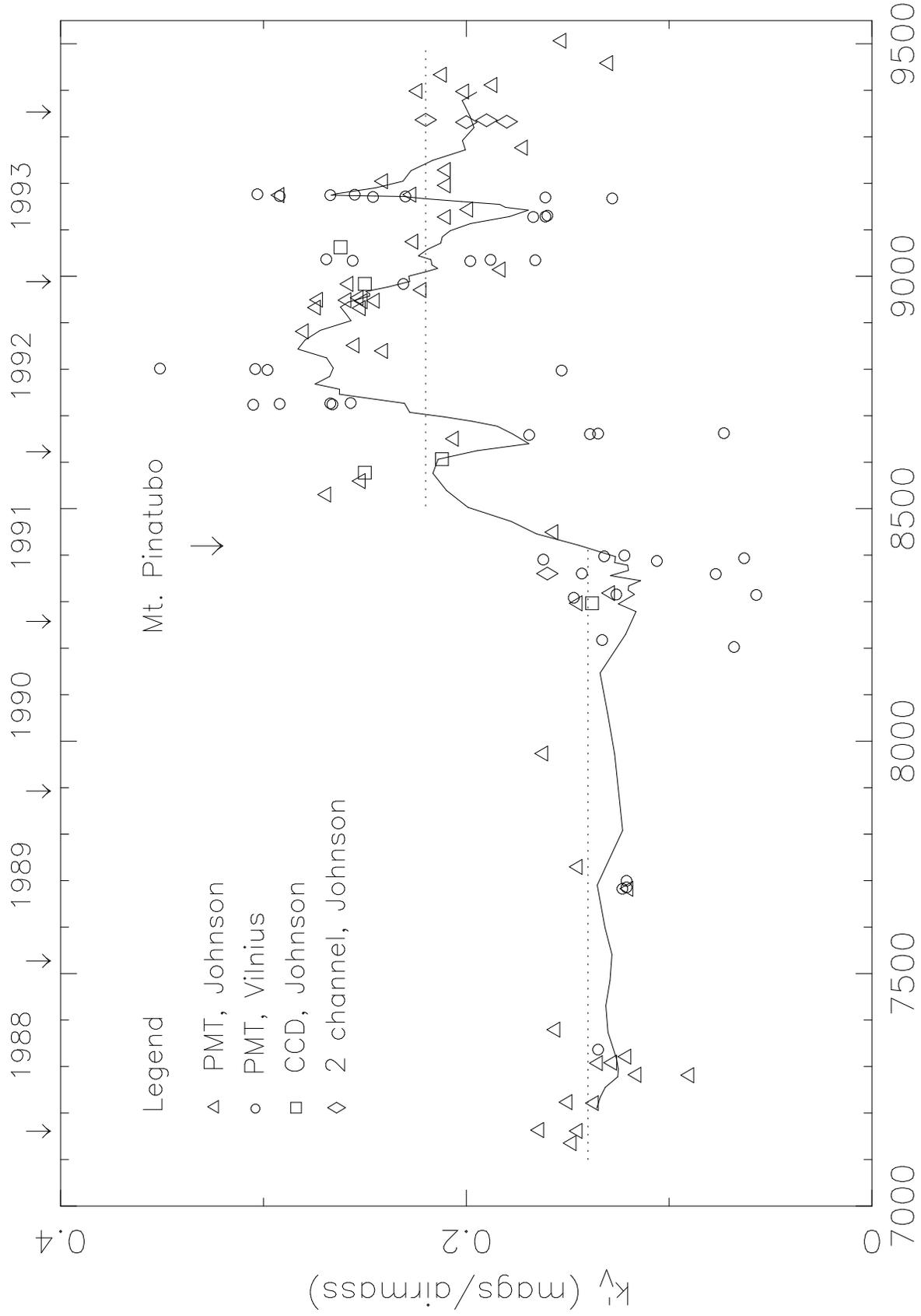
\end{document}